\documentclass[aip,apl]{revtex4-1}

\usepackage{graphicx}
\usepackage{dcolumn}
\usepackage{bm}

\begin{document}
\renewcommand{\thesection}{\arabic{section}} 
\title{The observation of electron trap liberation in MgF$_{2}$ doped with Yb$^{2+}$ using a two-color excitation experiment \\ }

\author{P. S. Senanayake}
\affiliation{Department of Physics and Astronomy, University of Canterbury, PB4800, Christchurch 8020, New Zealand}
\author{J.-P. R. Wells}

\email[Corresponding author:]{  jon-paul.wells@canterbury.ac.nz}
\affiliation{Department of Physics and Astronomy, University of Canterbury, PB4800, Christchurch 8020, New Zealand}
\author{M. F. Reid}
\affiliation{Department of Physics and Astronomy and MacDiarmid
  Institute for Advanced Materials and Nanotechnology, University of
  Canterbury, PB 4800, Christchurch 8140, New Zealand}
\author{G. Berden}
\affiliation{FELIX Free-Electron Laser Facility, FOM Institute for Plasmaphysics `Rijnhuizen', P.O. Box 1207, 3430 BE, Nieuwegein, The Netherlands}
\author{A. Meijerink}
\affiliation{Debye Institute for NanoMaterials Science, University of Utrecht, P.O. Box 80000, TA 3500, Utrecht, The Netherlands}
\author{R. J. Reeves}
\affiliation{Department of Physics and Astronomy and MacDiarmid
  Institute for Advanced Materials and Nanotechnology, University of
  Canterbury, PB 4800, Christchurch 8140, New Zealand}
\date{\today}

\begin{abstract}
\vspace{0.3cm}
\noindent
We utilize the optical transitions of Yb$^{2+}$ excited by an ultraviolet
optical parametric amplifier to probe electron trap liberation in
MgF$_{2}$ via the observation of a photoluminescence enhancement effect
induced by a subsequent infrared pulse from a free-electron laser. The
temperature dependence of the enhancement suggests that we liberate very
shallow traps having a depth of approximately 17 cm$^{-1}$. The
observed `trap spectrum' is consistent with a simple model of a Coulomb
trap.
\end{abstract}

\maketitle

The interconfigurational $f \rightarrow d$ transitions of both trivalent and divalent lanthanide ions have attracted extensive interest over recent years as increasing availability of high energy light sources have made comprehensive investigations of the energy-level structure of the excited configuration more readily practicable.\cite{ref1} These investigations are most often driven by possible applications in tunable UV lasers or efficient mercury-free lighting.

A complicating feature of the emission spectra obtained from some
lanthanide-doped materials is the presence of impurity-trapped excitons
(ITE) which form the lowest-energy emitting state under certain
conditions.\cite{ref2} The ITE consists of a hole localized on the
lanthanide ion and an electron delocalized on the neighboring metal
cations. The ITE emission is typified by a significant red shift and
broadening relative to the absorption, as well as long (up to tens of
milliseconds) lifetimes. ITE emission (often termed `anomalous' emission) is most commonly observed for divalent lanthanides (Eu$^{2+}$
and Yb$^{2+}$) residing on large lattice sites.

We have recently demonstrated that it is possible to study in detail the
energy level structure of ITE in CaF$_{2}$:Yb$^{2+}$ using a two-color
transient photoluminescence enhancement technique.\cite{ref3,ref4} This
technique uses short-pulsed UV excitation of the interconfigurational $f
\rightarrow d$ transitions with subsequent formation of the ITE. The pulsed
infrared output of a free-electron laser (FEL) is spatially overlapped
with the UV beam and given a short temporal delay relative to the
lifetime of the ITE emitting state(s) i.e.\ around 100 $\mu$s. As the FEL
wavelength is widely tunable we may scan through any intra-excitonic
transitions in a given wavelength region. When the FEL is resonant with
an ITE transition an increase in the luminescence intensity is observed
at the time delay given to the FEL pulse. This occurs because the first
excited exciton level (at 40 cm$^{-1}$) has a shorter radiative lifetime
than the lowest emitting level. Thus whenever we populate high-lying
exciton states, non-radiative relaxation leads to a significant
population in the first excited level yielding an increase in the
emission on a timescale associated with the overall lifetime of the
first excited state. 
In those experiments IR enhancement was also observed that decayed with
the radiative lifetime of the lowest ITE state. This was interpreted as liberation of electrons from shallow traps that were subsequently captured at the lanthanide sites.\cite{ref4}

Here we report results for MgF$_{2}$ crystals doped with divalent
ytterbium. The luminescent properties of MgF$_{2}$:Yb$^{2+}$ have been
reported previously.\cite{ref5} In MgF$_{2}$, divalent ytterbium
occupies the Mg$^{2+}$ site, which is six-coordinate and has D$_{\rm 2h}$ point group
symmetry. Characteristic $f \rightarrow d$ absorption and emission
appears to have been observed although it has been suggested that this
system could exhibit ITE emission.\cite{ref2} That the observed
emission has its origin in $d \rightarrow f$ transitions is fairly
strongly supported by the fact that the red shift between the lowest
energy absorption peaks and the emission is small and that vibronic
structure is observed. This is most certainly not the case in the
Yb$^{2+}$ doped CaF$_2$ and SrF$_2$ systems.

The experiments described here employed highly polished crystals cut from the same boule as those used in the earlier study of Lizzo.\cite{ref5} They contained 0.6 mol $\%$ of ytterbium and were grown using the Bridgman technique in a reducing atmosphere. 335 nm excitation was provided by a Quantronix TOPAS traveling wave optical parametric amplifier (OPA) having a repetition rate of 1 kHz and a pulse length of 3 ps. The Dutch free-electron laser FELIX in Nieuwegein was used to excite the samples in the infrared. The output of FELIX consists of a 6 $\mu$s long macropulse with a repetition rate of 10 Hz in the experiments performed here. The IR radiation was tuned between 5 and 25 $\mu$m. The two lasers were synchronized and have an electronically-variable
delay, which was limited to less than one millisecond due to the repetition rate of the OPA. The two beams were spatially overlapped on the sample with a spot size of close to 100 $\mu$m. The resultant fluorescence was collected using a TRIAX 320 spectrometer and detected with an RCA C31034 photomultiplier tube.

\begin{figure}[htb]
\centering
\includegraphics[width=0.6\columnwidth]{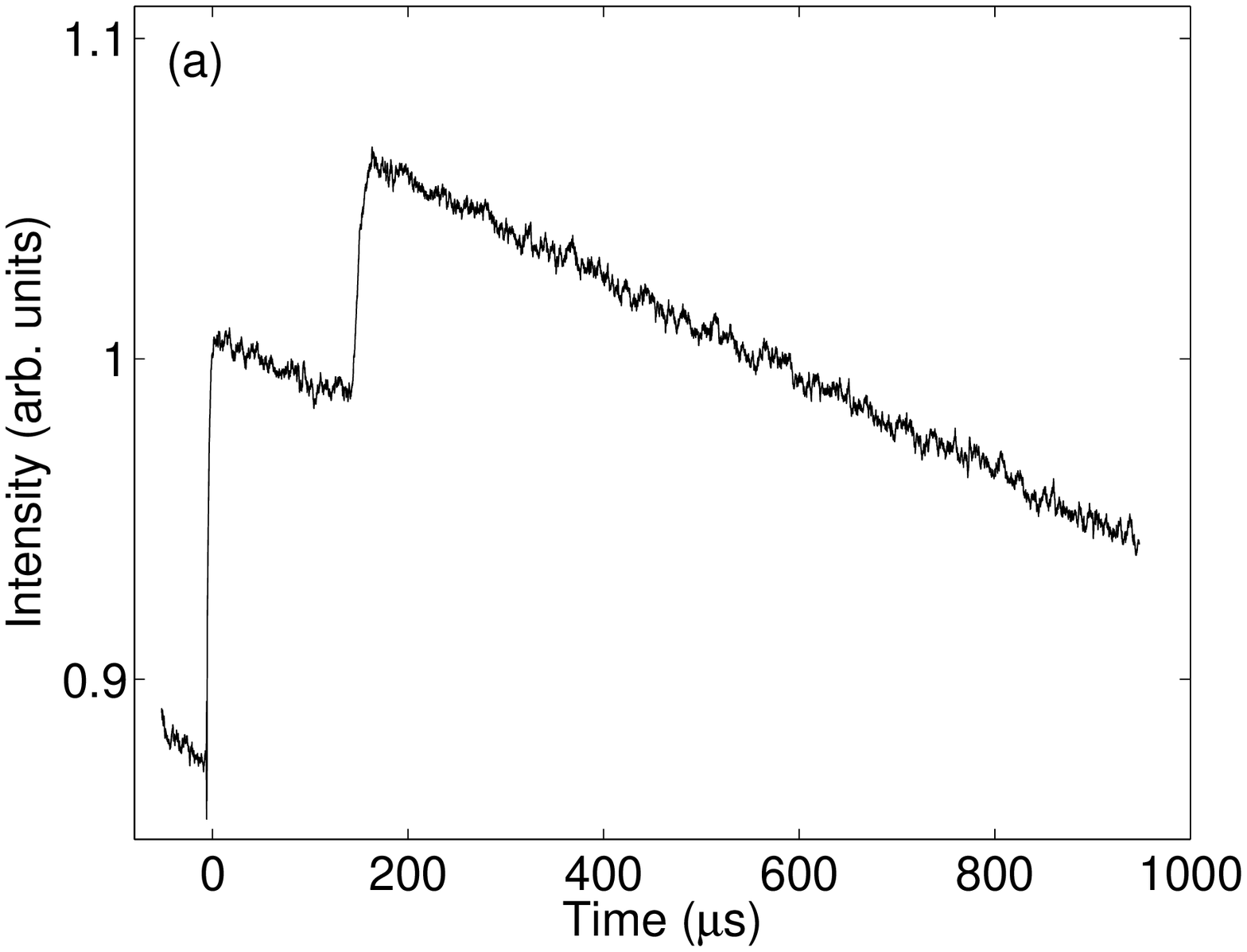}\\
\includegraphics[width=0.6\columnwidth]{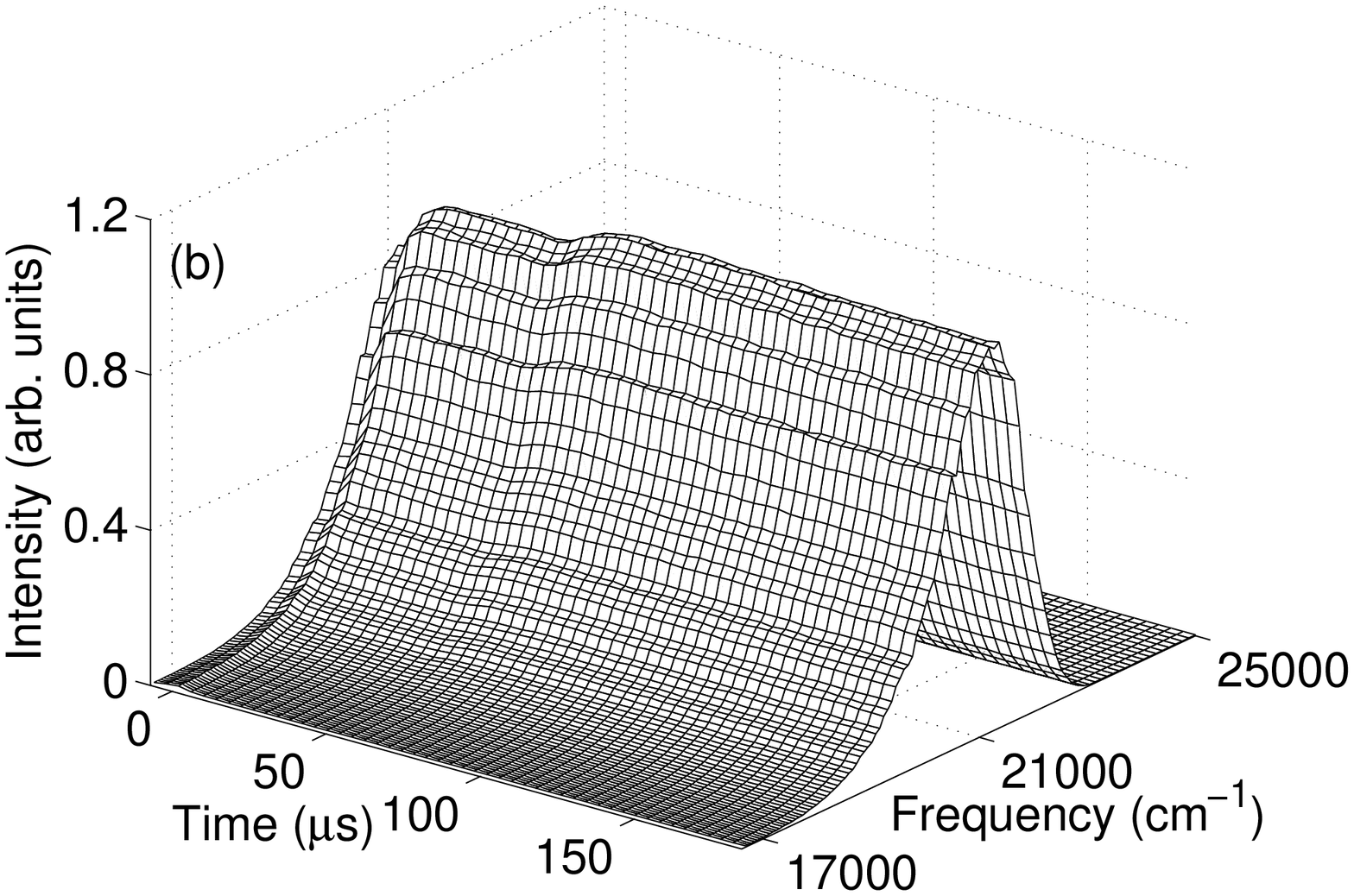}
\caption{\label{fig1}(a) 10~K photoluminescence decay for MgF$_{2}$:Yb$^{2+}$ showing enhancement under 150 $\mu$s delayed FEL excitation at 13.3 $\mu$m and (b) spectrally-resolved photoluminescent decay shown on a shorter time scale with the FEL excitation time delayed by 50 $\mu$s.}
\end{figure}

Figure \ref{fig1}(a) shows the 10~K photoluminescence decay of
MgF$_{2}$:Yb$^{2+}$ under UV excitation of the $f \rightarrow d$
inter-configurational transitions at a wavelength of 335 nm, through the
direct
excitation of the E band of Lizzo.\cite{ref5} Emission is monitored at
475 nm, which is the peak of the emission band. This underlying transient
is well approximated by a single exponential decay having a decay time
of 7.5 ms and is entirely consistent with the findings of reference \onlinecite{ref5}. The increase in photoluminescence observed at a time delay of 150 $\mu$s arises from the application of an infrared pulse from the FEL. The IR wavelength used was 750 cm$^{-1}$ 
(13.3 $\mu$m). It can be seen that the IR pulse induces an increase in
the emission detected from the sample of up to 8$\%$ and that the
enhanced emission component decays with the same characteristic lifetime
that is observed in the absence of the IR excitation. This suggests that
the enhanced emission arises from the same emitting state and is therefore quite unlike that observed for the ITE emission in CaF$_{2}$:Yb$^{2+}$ for example. We have also varied the time delay between the excitation pulses from 40 to 600 $\mu$s and have found no variation in the temporal characteristics. 

Figure \ref{fig1}(b) shows the temporal evolution of the emission spectrum under the application of both the UV and IR pulses with a shorter time delay of 50 $\mu$s between them. There is signal preceding the arrival of the UV pulse at $t=0$ because the lifetime of the Yb$^{2+}$ emission exceeds the repetition rate of UV laser. What is 
\begin{figure}[htb]
\centering
\includegraphics[width=0.6\columnwidth]{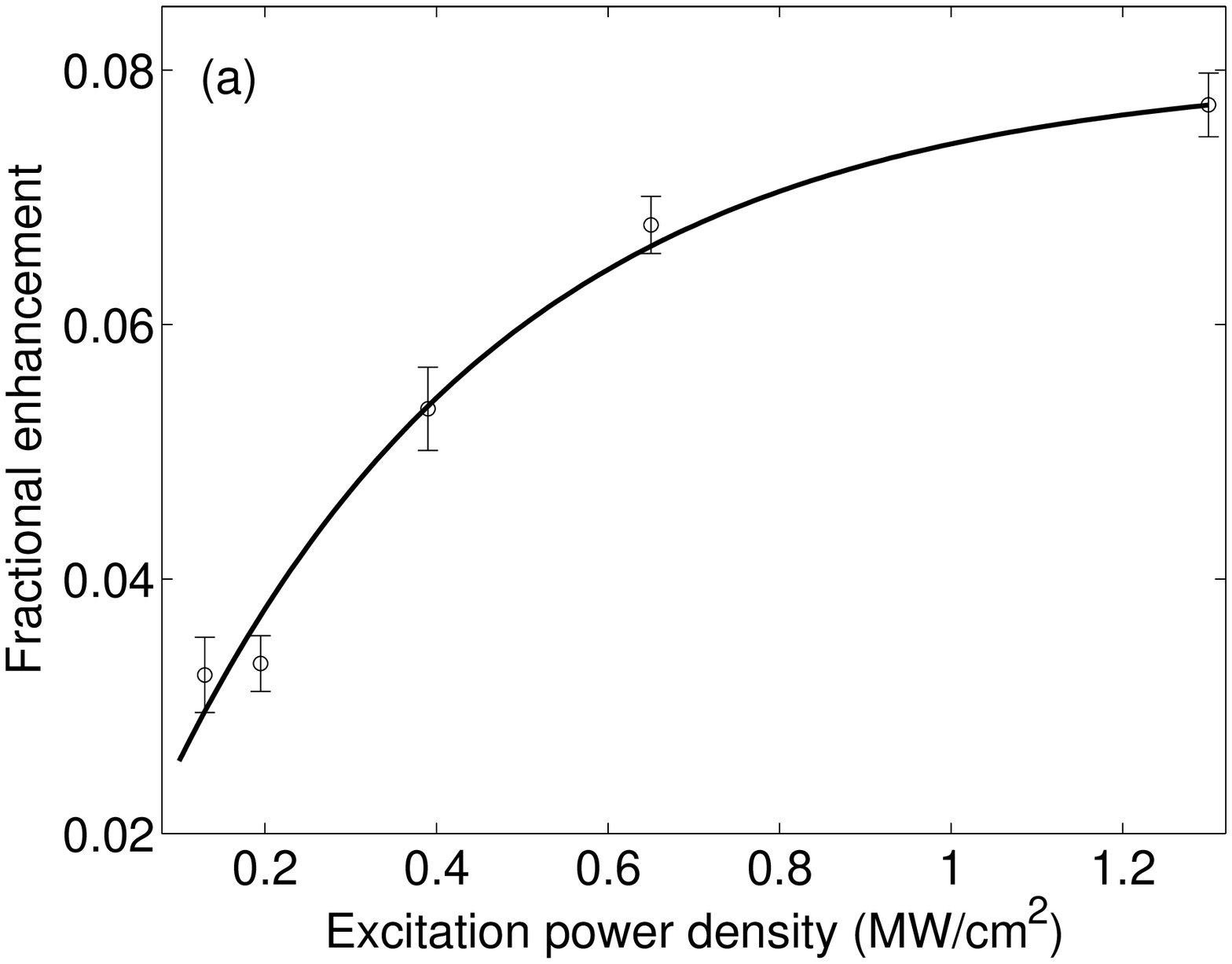}\\
\includegraphics[width=0.6\columnwidth]{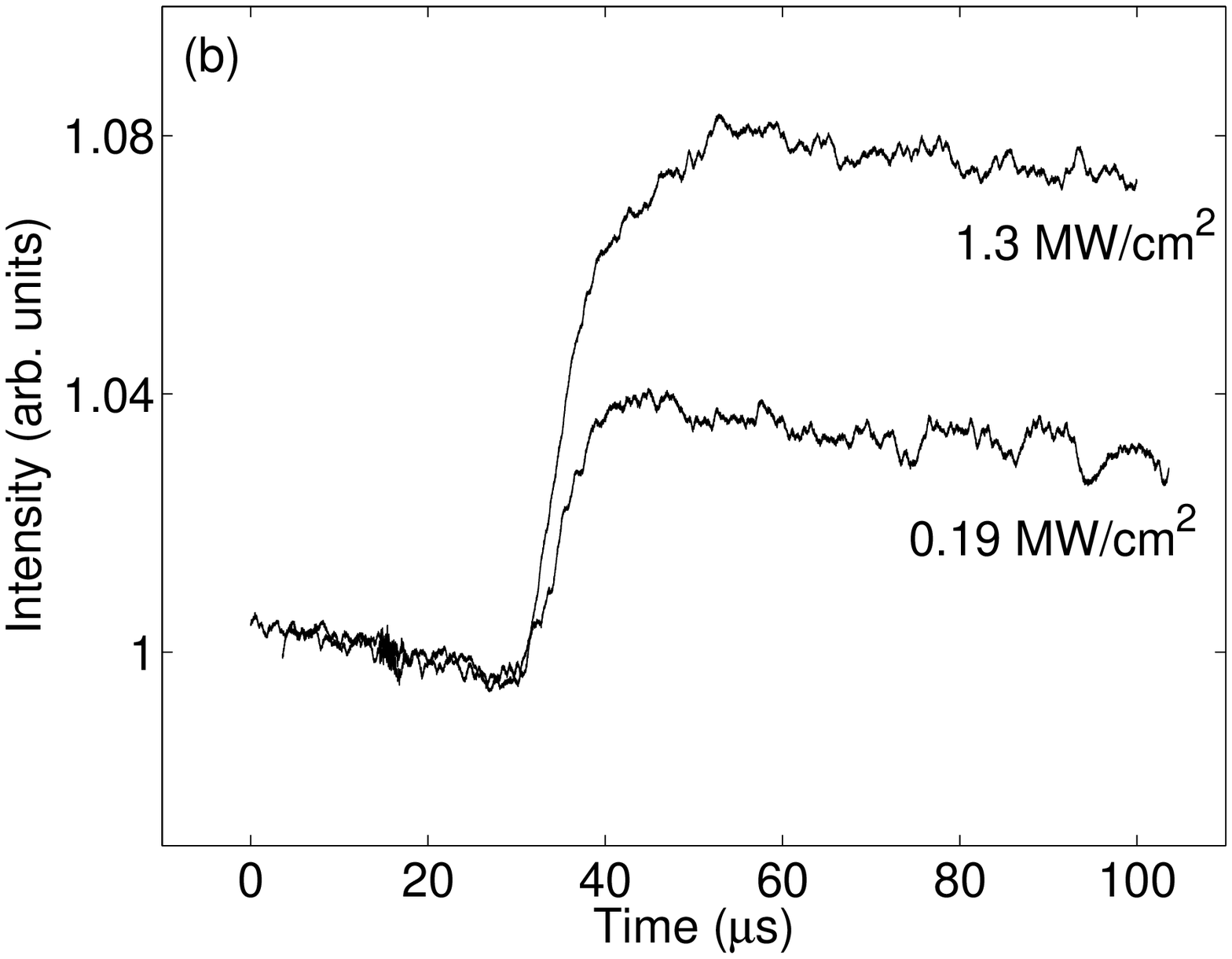}
\caption{\label{fig2}(a) Fractional magnitude of the enhancement effect as a
  function of IR power and (b) low- and high-power risetimes on the IR
  induced enhancement of the Yb$^{2+}$ photoluminescence. The solid line in (a) is a fit to an exponentially saturating process.\\}
\end{figure}
immediately obvious is that there is no spectral shift observed in the
spectrum upon the arrival of the IR pulse. Again, this is in contrast to
the spectral behavior of the ITE photoluminescence enhancement
(reference \onlinecite{ref3}, Figure \ref{fig1}(b)) where emission from
different states with substantially different bond lengths yields
considerable spectral shifts under IR excitation and therefore the
available population of emitters is shifted around in both the time and
spectral domains. Thus it seems reasonable to conclude that the observed
emission spectrum from  MgF$_{2}$:Yb$^{2+}$  is indeed $d \rightarrow f$ emission, as suggested in reference \onlinecite{ref5}.

We observe that for a given IR fluence the
fractional enhancement of the signal does not depend on the UV
excitation density up to 14 GW/cm$^2$, the maximum available from the
UV source.  However, the fractional enhancement increases with
IR excitation density, and there is clear evidence of the onset of saturation, as is shown in Figure \ref{fig2}(a). This is accompanied by an increasing rise-time observed on the transient photoluminescence enhancement itself (Figure \ref{fig2}(b)). The low-power rise-time essentially matches the macropulse length of 6 $\mu$s, i.e.\ it is instrument limited. However at the highest powers we observe rise-times as long as 21 $\mu$s.
\begin{figure}[htb]
\centering
\includegraphics[width=0.6\columnwidth]{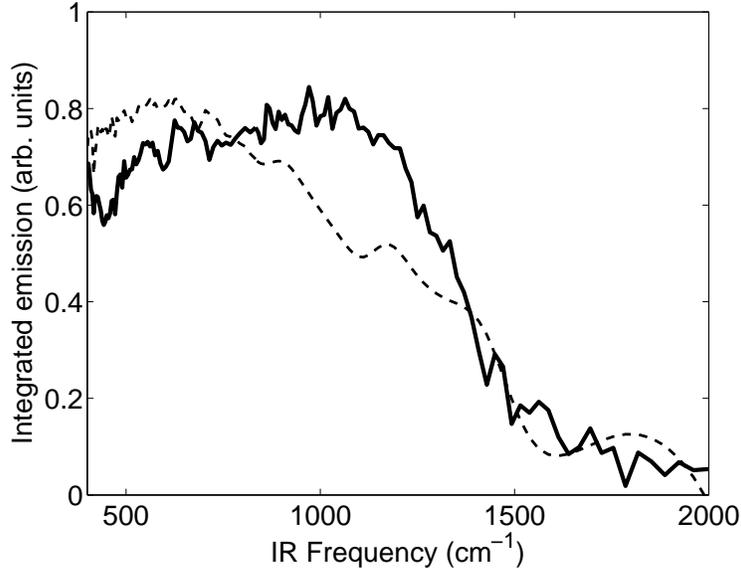}

\caption{\label{fig3}The solid line is the 10~K trap-liberation spectrum of MgF$_{2}$:Yb$^{2+}$ and the dashed line is the FEL power curve measured just before the cryostat.}
\end{figure}

The origin of the photoluminescence enhancement can most likely be attributed to the IR pulse liberating electron traps, which become populated upon application of the UV pulse through photoionization into the conduction band. In effect, the IR pulse at least partially reverses the cycle and the liberated electrons relax back into the Yb$^{2+}$ $d$ shell. Therefore the IR pulse increases the total available population of emitters and an enhancement of the 
emission intensity can be observed. This interpretation explains several
observations. In particular the observed saturation of the transient
enhancement under increasing IR excitation which can be explained as the
available shallow trap states being depleted. Further the increasing
rise-time observed is likely associated with electrons occupying high-mobility states within the conduction band and reflects the time taken
for the excited electron to find a luminescent center.
The constant fractional enhancement with UV excitation power density shows that at least up to 14 GW/cm$^2$ we are not yet saturating the empty traps available to receive the photoionized electrons.

Production of traps in MgF$_2$  by  high-energy ionizing radiation or UV
excitation has been previously studied\cite{ref6, ref7} Many
of these traps have excited states lying just below the conduction band.
For example, the F center excited state is 565 cm$^{-1}$ below the conduction band \cite{ref6} and is an efficient intrinsic electron trap. One can estimate the depth of the traps we are observing using the temperature dependence of the signal. It is found that the enhancement effect diminishes in amplitude extremely quickly as the sample temperature increases above 10~K and is no longer detectable above a temperature of 20~K. This occurs due to thermal escape of electrons and we determine a trap depth of approximately 17 cm$^{-1}$. 

Figure \ref{fig3} shows the spectral dependence of the transient
enhancement. The signal has been integrated over time, and the underlying luminescence from the
UV excitation alone subtracted. 
The spectrum is essentially flat between 400 and 1100 cm$^{-1}$ whilst
the apparent drop in signal at higher energies matches very well the FEL
power, which reduces at shorter wavelengths for the laser settings used
in this experiment. A simple model of a Coulomb or delta trap\cite{ref8}
with a trap depth of 17 cm$^{-1}$ is consistent with our observed
spectral response, since in this experiment the excitation energy is in the essentially flat high-energy tail.

In summary, we have observed trap liberation in MgF$_{2}$:Yb$^{2+}$ in a two-color excitation experiment that uses infrared pulses to liberate trapped electrons subsequent to excitation of Yb$^{2+}$ at 335 nm. A risetime can be observed on the induced signal which is as long as 21 $\mu$s for high infrared excitation densities and the magnitude of the effect appears to saturate. The trap-liberation spectrum is essentially flat over the wavelength range studied and this is consistent with the observed temperature dependence of the effect from which we derive a trap depth of 17 cm$^{-1}$ under the assumption of a Coulomb trap model. This value is quite different from CaF$_{2}$ and SrF$_{2}$ \cite{ref3,ref4} which have trap depths around 400 cm$^{-1}$. However MgF$_{2}$ has the rutile structure and is, therefore, fundamentally different from both CaF$_{2}$ and SrF$_{2}$.

\bigskip

The authors acknowledge support from the Marsden fund of the Royal Society of New Zealand under contract 09-UOC-080. We thank the Dutch FOM organization
for providing the required FELIX beamtime and the FELIX staff for their
assistance. P.S.S. acknowledges support from the University of
Canterbury.

\end{document}